# Formation of epitaxial *p-i-n* structures on the basis of (In,Fe)Sb and (Ga,Fe)Sb layers

A.V. Kudrin[1,*], V.P. Lesnikov[1], D.A. Pavlov[1], Yu.V. Usov[1], Yu.A. Danilov[1], M.V. Dorokhin[1], O.V. Vikhrova[1], V.E. Milin[1], R.N. Kriukov[1], Yu.M. Kuznetsov[1], V.N. Trushin[1], and N.A. Sobolev[2,3]

[1]*Lobachevsky State University of Nizhny Novgorod, Gagarin av. 23/3, 603950 Nizhny Novgorod, Russia*
[2]*Department of Physics and I3N, University of Aveiro, 3810-193 Aveiro, Portugal*
[3]*National University of Science and Technology "MISiS", 119049 Moscow, Russia*
*kudrin@nifti.unn.ru

Multilayer structures on the basis of *n*-type (In,Fe)Sb and *p*-type (Ga,Fe)Sb diluted magnetic semiconductors (DMS) along with separate (In,Fe)Sb and (Ga,Fe)Sb layers were fabricated on GaAs substrates by pulsed laser sputtering of InSb, GaAs, GaSb, Sb and Fe targets in a vacuum. Transmission electron microscopy and energy-dispersive x-ray spectroscopy reveal a strong dependence of the phase composition of the (In,Fe)Sb compound on the growth temperature. An increase of the latter from 220ºC to 300ºC leads to a coalescence of Fe atoms and formation of a secondary crystalline phase in the (In,Fe)Sb layer with a total Fe content of $\approx 10$ at. %. At the same time, the $Ga_{0.8}Fe_{0.2}Sb$ layers obtained at 220ºC and 300ºC are single-phase. The separate $In_{0.8}Fe_{0.2}Sb$ and $Ga_{0.8}Fe_{0.2}Sb$ layers grown on *i*-GaAs at 220ºC are DMS with Curie temperatures of $\approx 190$ K and 170 K, respectively. The three-layer *p-i-n* diode (In,Fe)Sb/GaAs/(Ga,Fe)Sb structure grown on a GaAs substrate at 220ºC with a Fe content of $10 \pm 1$ at. % in the single-phase (In,Fe)Sb and (Ga,Fe)Sb layers has a rather high crystalline quality and can be considered as a prototype of a bipolar spintronic device based on Fe-doped III-V semiconductors.

## I. INTRODUCTION

Presently the formation of III-V semiconductor layers heavily doped with Fe atoms, (in particular (In,Fe)As [1,2], (Ga,Fe)Sb [3-5], (In,Fe)Sb [6-9], (Al,Fe)Sb [10]) is a new and promising trend in the field of semiconductor spintronics. The (Ga,Fe)Sb and (In,Fe)Sb layers fabricated by molecular beam epitaxy [4,7,8] and pulsed laser sputtering [5,6,9] demonstrate intrinsic ferromagnetism up to room temperature. At present, the electrical activity of iron in the InSb and GaSb matrices is not completely clear. According to the data available, the Fe impurity in the InSb and GaSb hosts forms a level within the valence band. This should exclude the manifestation of donor or acceptor properties of the Fe dopant in these semiconductor matrices [3,6]. However, investigations reveal the *n*-type conductivity in (In,Fe)Sb layers [6-9] and *p*-type conductivity in (Ga,Fe)Sb layers [3-5]. This occurs due to electrically active structure defects in these semiconductor crystals, the presence of which leads to the formation of *n*-type InSb layers [11] and *p*-type GaSb layers [12] without the introduction of electrically active dopants. Thus, on the basis of the (In,Fe)Sb and (Ga,Fe)Sb diluted magnetic semiconductors (DMS) a prototype of a room temperature bipolar spintronics device in form of a multilayer heterostructure can be fabricated. For the fabrication of *p-n* or *p-i-n* diode structures, it is necessary to develop a technology of epitaxial growth on the surface of heavily Fe-doped III-V magnetic semiconductors. In particular, undoped GaAs is a suitable material for the *i*-region of a *p-i-n* structure. In addition, GaAs is a suitable material as a substrate as well. One of the main problems of the formation in a unique technological cycle of multilayer structures comprising (In,Fe)Sb, (Ga,Fe)Sb and GaAs semiconductors is a large lattice mismatch which amounts to 14.6% in the InSb/GaAs pair and to 7.8% in the GaSb/GaAs pair.

In this work we present the results of a study of multilayer epitaxial heterostructures containing layers of the (In,Fe)Sb and (Ga,Fe)Sb diluted magnetic semiconductors, and also of separate (In,Fe)Sb and (Ga,Fe)Sb layers.

## II. EXPERIMENTS

Samples were grown by pulsed laser sputtering of InSb, GaAs, GaSb, Sb and Fe targets in a vacuum chamber with a background gas pressure of about $2 \times 10^{-5}$ Pa. The presence of an additional Sb target allowed us to introduce an additional amount of antimony during the sputtering process. The technological Fe content was set by the growth parameter $Y_{Fe} = t_{Fe}/(t_{Fe}+t_{InSb(GaSb)})$, where $t_{Fe}$ and $t_{InSb/GaSb}$ are the ablation times of the Fe and InSb (or GaSb) targets, respectively. The $Y_{Fe}$ value in all (In,Fe)Sb and (Ga,Fe)Sb layers was $Y_{Fe} = 0.17$.

Four samples named A, B, C and D were grown.

Sample A contains the following sequence of layers: GaAs buffer layer (the planned technological thickness ($d_{tech}$) is ~ 20 nm), (In,Fe)Sb layer ($d_{tech}$ ~ 40 nm), GaAs spacer layer ($d_{tech}$ ~ 50 nm), (Ga,Fe)Sb layer ($d_{tech}$ ~ 40 nm), and GaAs cap layer ($d_{tech}$ ~ 15 nm). The GaAs layers were undoped. The growth temperature was of 300ºC. Semi-insulating (001) GaAs was used as a substrate.

Sample B comprises the following set of layers: (In,Fe)Sb layer ($d_{tech}$ ~ 20 nm), undoped GaAs spacer layer ($d_{tech}$ ~ 20 nm), (Ga,Fe)Sb layer ($d_{tech}$ ~ 20 nm). The growth temperature was of 220ºC. Schematic drawings of samples A and B are shown in Figure 1(a). *n*-Type (001) GaAs was used as a substrate.

Sample C and sample D are single (In,Fe)Sb and (Ga,Fe)Sb layers, respectively, with $d_{tech}$ ~ 50

nm, grown at 220ºC on semi-insulating (001) GaAs substrates.

During the (In,Fe)Sb layers formation, the introduction of an additional amount of antimony from the Sb target was carried out for the suppression of the indium islands formation on the surface as a consequence of the Sb deficiency [6]. Our previous studies revealed that the formation of a relatively smooth (Ga,Fe)Sb layer by the used technique is possible without the introduction of an additional amount of antimony [5].

The structural properties of the samples were investigated by x-ray diffraction (XRD) and high-resolution analytical transmission electron microscopy (HRA TEM). The distribution of the constituent elements was obtained by energy-dispersive x-ray spectroscopy during microscopy investigations. Optical reflectivity spectra of samples C and D were obtained at room temperature in the spectral range from 1.6 – 6 eV. DC magnetotransport measurements were carried out in the temperature range from 10 – 300 K in a closed-cycle He cryostat.

## III. RESULTS AND DISCUSSION

### A. Multilayer structures

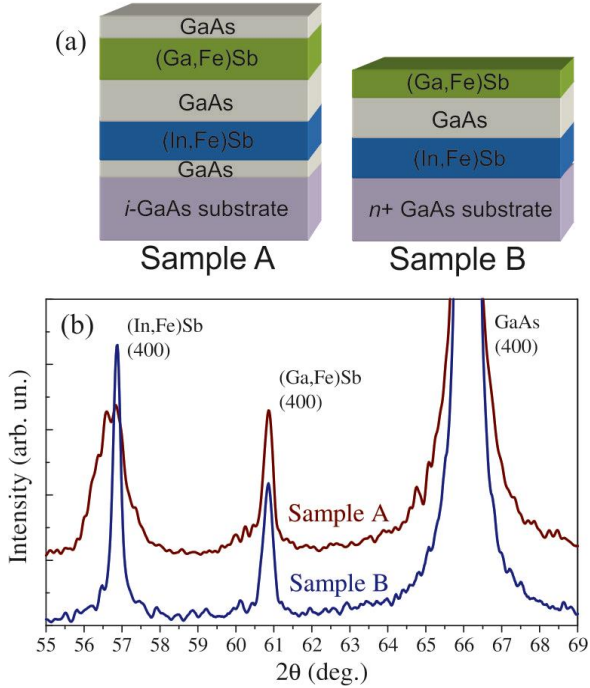

FIG. 1. (a) Schematic drawings of samples A and B. (b) X-ray diffraction curves for samples A and B.

Figure 1(b) shows $\theta$-$2\theta$ XRD scans for samples A and B measured at room temperature using Cu K$_{\alpha1}$ radiation. The XRD curves contain three reflections: the high-intensity (400) reflection from the GaAs substrate (overlapped with the reflection from the GaAs spacer layer) and the (400) reflections from the (In,Fe)Sb and (Ga,Fe)Sb layers. The $\theta$-$2\theta$ scans reveal the epitaxial growth character of all layers in both samples.

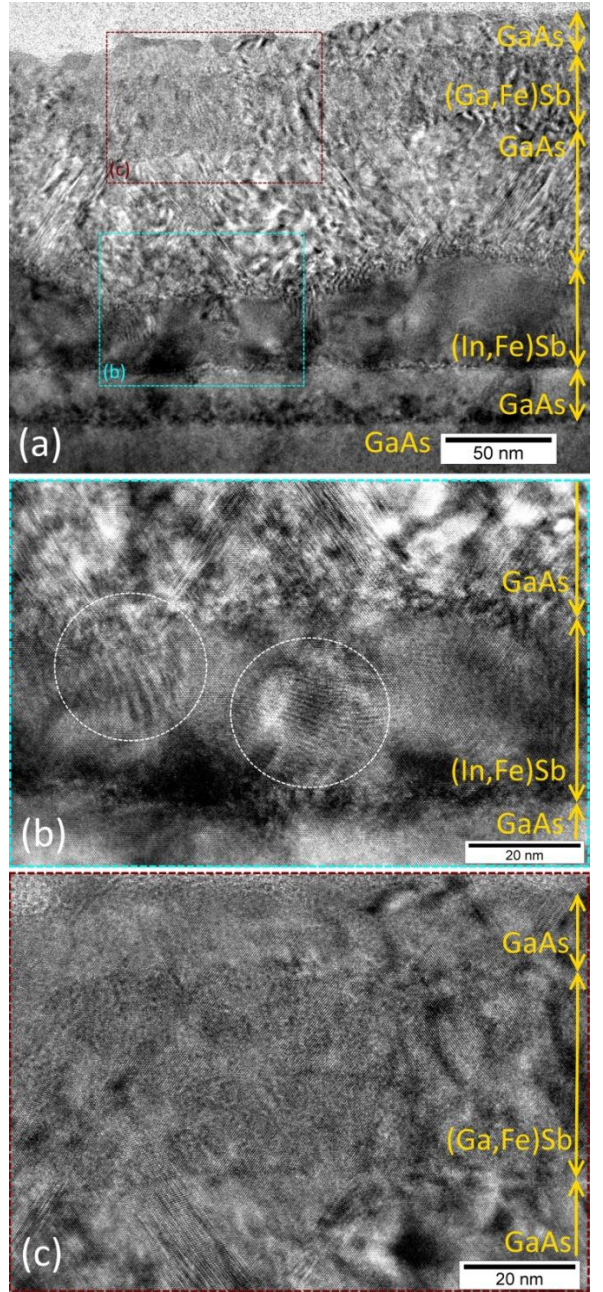

FIG. 2. (a) Overview TEM image of sample A. (b) HR TEM image of the GaAs buffer/(In,Fe)Sb/GaAs spacer region (representing the enlarged area of the dashed square (b) in [Figure 2(a)]). (c) HR TEM image of the GaAs spacer/(Ga,Fe)Sb/GaAs cap region (the enlarged area of the dashed square (c) in [Figure 2(a)]).

Figure 2(a) shows an overview cross-section TEM image of sample A. The image was obtained from a 270 nm long region. The image clearly reveals the grown layers: GaAs buffer layer (≈ 25 nm thick), (In,Fe)Sb layer (with varying thickness in the range from 30 – 50 nm), GaAs spacer layer (≈ 60 nm thick), (Ga,Fe)Sb layer (≈ 35 nm thick), and GaAs cap layer (≈ 17 nm thick). It is noteworthy that there is a relatively strong thickness variation of the (In,Fe)Sb layer. Figure 2(b) shows a high-resolution TEM (HR TEM) image of a 105 nm long region containing a part of the GaAs buffer, the (In,Fe)Sb layer and a part of the GaAs spacer. The image reveals areas with a moiré-like contrast within the (In,Fe)Sb layer (the area in dashed circles in [Figure 2(b)]). These areas are Fe-enriched regions, i.e. secondary-phase inclusions. The presence of these inclusions in the (In,Fe)Sb layer and its relatively significant roughness indicate that the growth temperature of 300ºC is not optimal. Despite the roughness and multi-phase character of the (In,Fe)Sb layer, the growth of subsequent layers is epitaxial. As can be seen in Figure 2(b), the atomic rows in the GaAs spacer are a continuation of the atomic rows in the (In,Fe)Sb layer. The GaAs spacer contains a large number of stacking faults in the {111} planes, that appear in the image as an assemblage of straight lines at an angle of ≈ 70º with respect to each other [Figure 2(b)]. The stacking faults generation is a consequence of the large lattice mismatch between the (In,Fe)Sb and GaAs matrixes [6]. Figure 2(c) shows a HR TEM image of a 105 nm long region containing a part of the GaAs spacer, the (Ga,Fe)Sb layer and the GaAs cap layer. Both layers, (Ga,Fe)Sb and the GaAs cap, are epitaxial. The continuation of crystal rows from the GaAs spacer layer to the surface is well seen in Figure 2(c). Contrary to the (In,Fe)Sb layer, the formation of a secondary crystalline phase in the (Ga,Fe)Sb layer is not observed, probably due to a better stability of (Ga,Fe)Sb with respect to the formation of a Fe-enriched secondary phase. The HR TEM images reveal that the (In,Fe)Sb, (Ga,Fe)Sb and GaAs spacer layers are fully relaxed. The lattice mismatch between the GaAs buffer and the (In,Fe)Sb layer and also between the (In,Fe)Sb layer and the GaAs spacer in the growth direction and in the plane (detected from HR TEM images) equals ≈ 14%. The lattice mismatch between the GaAs spacer and the (Ga,Fe)Sb layer in the growth direction and in the plane amounts to ≈ 8%. These values coincide with the $\Delta a/a$ values for the GaAs/InSb and GaAs/GaSb pairs. The geometric phase analysis allows one to estimate the relaxation layer thickness (the layer within which the transition to the maximum lattice mismatch occurs). For the GaAs (buffer)/(In,Fe)Sb heterojunction the relaxation layer thickness within the (In,Fe)Sb layer is about 3 nm. For the (In,Fe)Sb/GaAs (spacer) heterojunction the relaxation layer within the GaAs spacer is about 9 nm thick. For the GaAs (spacer)/(Ga,Fe)Sb heterojunction the relaxation layer thickness within the (Ga,Fe)Sb layer amounts to about 4 nm.

Figure 3 presents a scanning cross-section TEM (STEM) image of sample A and an energy-dispersive x-ray spectroscopy (EDS) mapping of Ga, As, In, Sb, and Fe constituent elements. The data reveal a non-uniform distribution of Fe atoms within the (In,Fe)Sb layer, which is consistent with the TEM studies [Figure 2(b)].

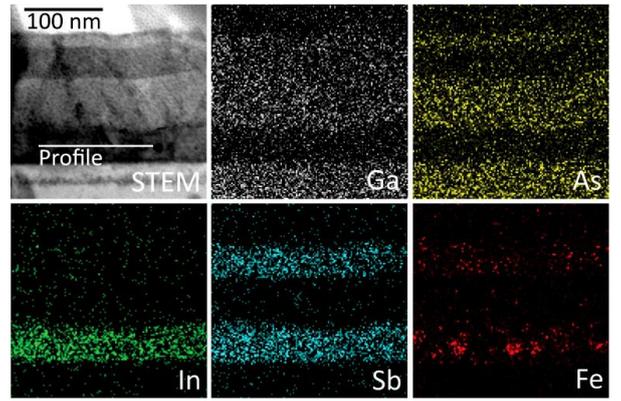

FIG. 3. STEM image and corresponding EDS mapping of the Ga, As, In, Sb and Fe constituent elements in sample A.

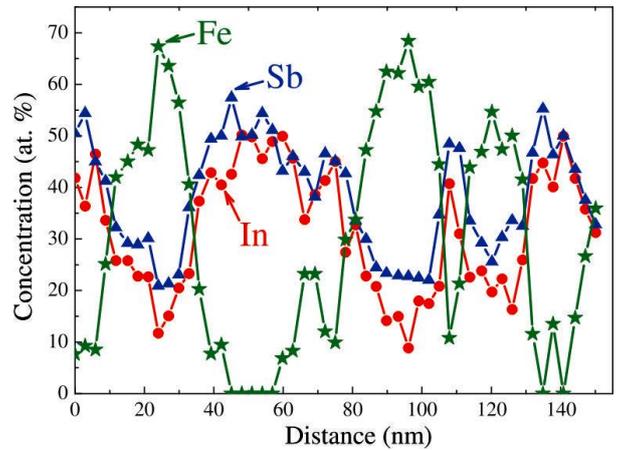

FIG. 4. EDS concentration profile of the In, Sb and Fe atoms in the (In,Fe)Sb layer of sample A.

Figure 4 shows the EDS concentration profile of the In, Sb, and Fe atoms in the (In,Fe)Sb layer of sample A along the line named "Profile" in the STEM image (Figure 4). The profile reveals that the Fe content in the Fe-enriched regions is up to 70 at. %. In the Fe-enriched regions a decrease in the In and Sb content is observed, whereas the concentration of the In and Sb atoms approximately coincides. The absence of detectable Fe atoms between the regions with high Fe concentrations is remarkable (Figure 4). According to the Springer Materials database [13], the crystalline chemical compounds $Fe_{1.17}Sb$, $FeSb_2$, $FeSb_3$ and compounds with a strong Fe predominance ($Fe_{0.97}Sb_{0.03}$, $Fe_{0.97}Sb_{0.03}$) are known. The chemical compounds of Fe and In were not reported in the literature. Taking into account the concentration profile (Figure 4), it is most probable that at $T_g$ = 300ºC the Fe atoms coalesce and form crystalline clusters of pure Fe (or $Fe_{0.97}Sb_{0.03}$/$Fe_{0.97}Sb_{0.03}$-like clusters with a strong predominance of Fe) in the InSb matrix. The cluster diameter is about 20 nm. The appearance of areas with a moiré-like contrast within the (In,Fe)Sb layer (Figure 2(b)) is a consequence of an overlap of the (In,Fe)Sb host and Fe cluster lattices. The average Fe content in the (Ga,Fe)Sb layer detected by EDS equals 10 ± 1 at. %.

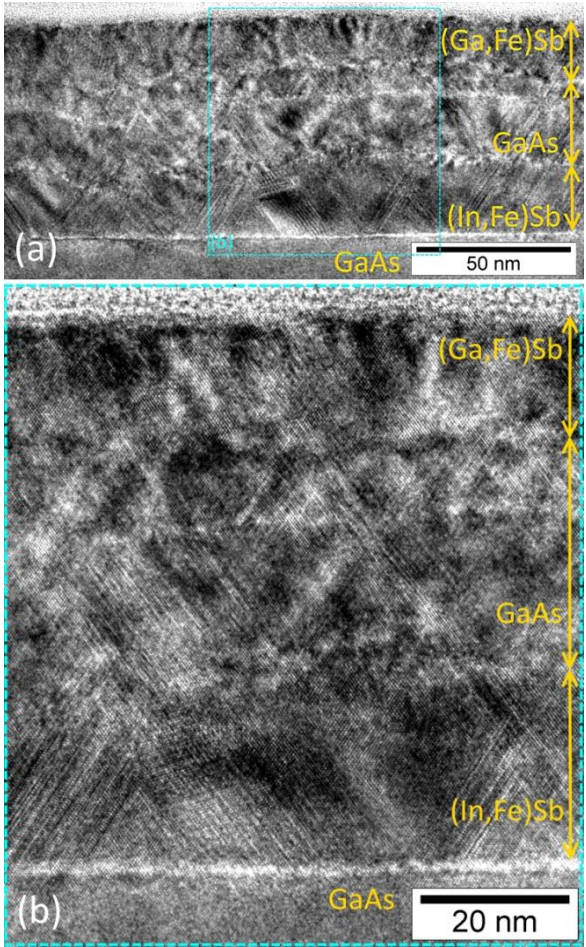

FIG. 5. (a) Overview TEM image of sample B. (b) HRTEM image of a 78 nm long region (the enlarged area in dashed square (b) on [Figure 5(a)]).

Figure 5(a) exhibits an overview cross-section TEM image of sample B. The image was obtained from a 190 nm long region. The image clearly reveals the following layers: (In,Fe)Sb layer (≈ 27 nm thick), GaAs spacer layer (≈ 27 nm thick), and (Ga,Fe)Sb layer (≈ 17 nm-thick). Figure 4(b) shows a HR TEM image of a 78 nm long region. Sample B demonstrates a significantly higher crystalline perfection as compared to sample A. The (In,Fe)Sb layer grown at the lower temperature ($T_g$ = 220ºC) with respect to structure A is rather smooth and does not contain regions with a moiré-type contrast, i.e. the layer is single-phase. In the (Ga,Fe)Sb layer the presence of a secondary crystalline phase was not detected either. The growth character of all layers is epitaxial. The preservation of atomic rows from the GaAs substrate to the surface is observed [Figure 5(b)]. Similarly to sample A, sample B contains a large number of stacking faults threading through all layers up to the surface. This, as was mentioned above, is a consequence of the large lattice mismatch between the (In,Fe)Sb and GaAs matrixes. The (In,Fe)Sb, (Ga,Fe)Sb and GaAs spacer layers, as in sample A, are completely relaxed. The values of the relaxation layer thickness are similar to the respective values in the corresponding heterojunctions of structure A.

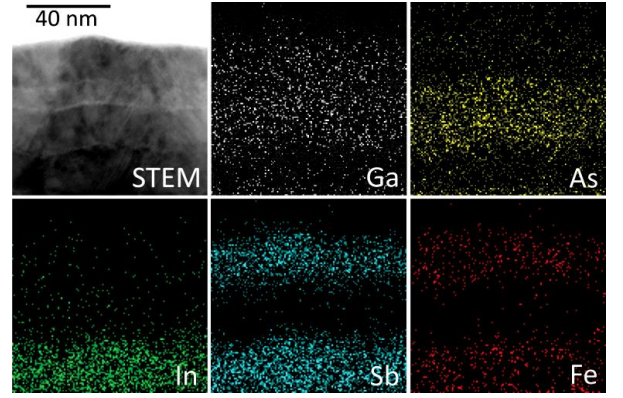

FIG. 6. STEM image and corresponding EDS mapping of Ga, As, In, Sb and Fe constituent elements in sample B.

Figure 6 shows a STEM image of sample B and EDS mapping of the Ga, As, In, Sb, and Fe constituent elements. In contrast to structure A (Figure 3), there are no obvious areas with a greatly increased Fe content. This, together with the HR TEM results, allows us to conclude about the relatively uniform distribution of Fe in the (In,Fe)Sb and (Ga,Fe)Sb layers without coalescence of the Fe atoms and cluster formation. The average Fe content in the (In,Fe)Sb and (Ga,Fe)Sb layers detected by EDS equals 10 ± 1 at. %.

For current-voltage (*I-V*) measurements of sample B, mesa structures with a diameter of 500 μm were prepared using a photolithography and chemical etching. The top Au/Ti and bottom Sn ohmic contacts were fabricated by electron beam evaporation and spark fusion of a tin foil, respectively. Figure 7 presents current-voltage characteristics of sample B measured at 300 K and 10 K. The *I-V* curves show that sample B functions as a diode.

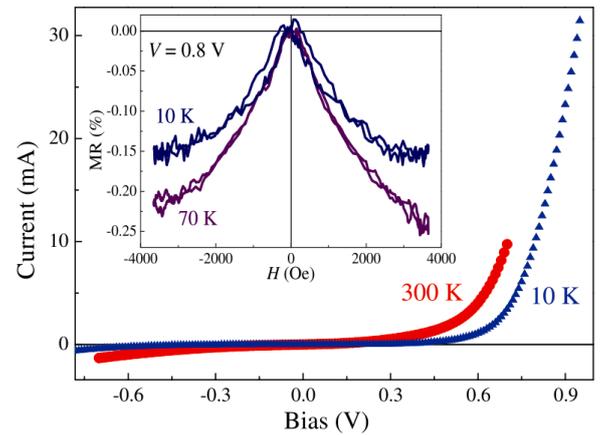

FIG. 7. *I-V* curves of sample B at 300 K and 77 K. The inset shows MR curves at 10 K and 70 K (*B* is applied perpendicular to the sample plane).

The inset of Figure 7 shows magnetoresistance curves (MR = (*R(B)* − *R*(0))/*R*(0), where *R* is the resistance of the structure at a given bias) taken at 10 K and 70 K with the external magnetic field applied perpendicular to the sample plane (parallel to the current flow). The hysteretic shape of the MR dependence at 10 K indicates that the magnetic layers affect the current transfer.

## B. (In,Fe)Sb/GaAs and (Ga,Fe)Sb/GaAs structures

For the investigation of the magnetic properties of the (In,Fe)Sb and (Ga,Fe)Sb layers included in structure B, structures C ((In,Fe)Sb/GaAs) and D ((Ga,Fe)Sb/GaAs) were obtained with the same technological parameters ($Y_{Fe} = 0.17$, $T_g = 220°C$).

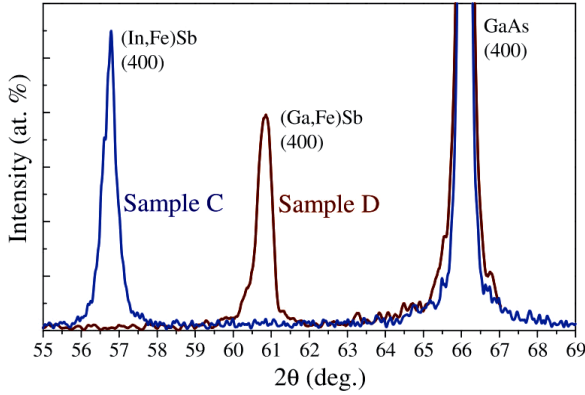

FIG. 8. XRD curves for the (In,Fe)Sb/GaAs (sample C), (Ga,Fe)Sb/GaAs (sample D) structures.

Figure 8 shows $\theta$-$2\theta$ XRD scans for samples C and D. In addition to the high-intensity GaAs peak, the XRD curves contain only (400) reflections from the (In,Fe)Sb and (Ga,Fe)Sb layers for structures C and D, respectively. The position and width of the peaks are close to the corresponding peaks from the (In,Fe)Sb and (Ga,Fe)Sb layers in the $\theta$-$2\theta$ XRD scan for structure B (Figure 1) and also reveal the epitaxial growth character of the (In,Fe)Sb and (Ga,Fe)Sb layers in structures C and D.

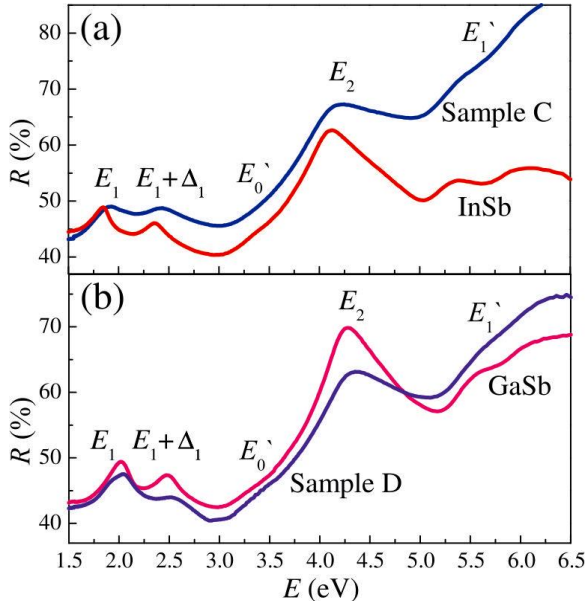

FIG. 9. Optical reflectivity spectra taken at 295 K on the (In,Fe)Sb/GaAs (sample C), (Ga,Fe)Sb/GaAs (sample D) structures and on InSb, GaSb bulk crystals.

Figure 9 exhibits optical reflectivity spectra measured at 295 K on the structures (In,Fe)Sb/GaAs (sample C), (Ga,Fe)Sb/GaAs (sample D) and on InSb, GaSb bulk crystals for comparison. The reflectivity spectra of the obtained single (In,Fe)Sb and (Ga,Fe)Sb layers agree with the spectra of the bulk InSb and GaSb crystals, respectively. The spectra contain peaks related to characteristic interband transitions, in particular a clear doublet in the $E_1$ region and an intense peak in the $E_2$ region [14]. Both for the (In,Fe)Sb and (Ga,Fe)Sb layers the peaks in the $E_1$ region exhibit a blueshift. A similar blueshift is characteristic of the (Ga,Fe)Sb [3] and (In,Fe)Sb compounds [7,9]. The reflectivity spectra verify a good crystalline quality of the (In,Fe)Sb and (Ga,Fe)Sb layers and display the conservation of the band structure of the InSb and GaSb semiconductor hosts.

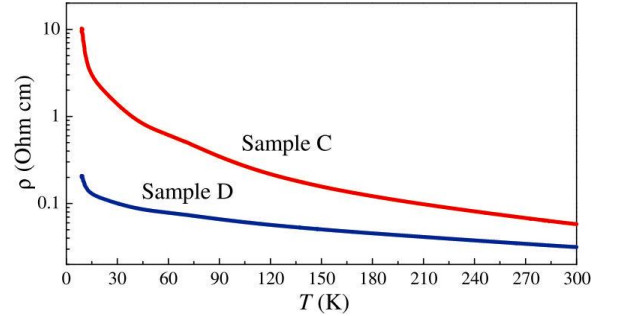

FIG. 10. Temperature dependences of the resistivity for the (In,Fe)Sb/GaAs (sample C) and (Ga,Fe)Sb/GaAs (sample D) structures.

Figure 10 shows the temperature dependences of the resistivity $\rho(T)$ for the (In,Fe)Sb and (Ga,Fe)Sb layers. The resistivity of the (In,Fe)Sb layer is significantly higher than that of the (Ga,Fe)Sb layer, especially at low temperatures. The higher resistivity of the (In,Fe)Sb layer is apparently related to a lower density of electrically active defects which are predominantly donors in the InSb [11] and acceptors in the GaSb [12] hosts. The Fe impurity does not show a clear electrical activity in the InSb and GaSb hosts, therefore, the conductivity type of the (In,Fe)Sb [6-9] and (Ga,Fe)Sb [3-5] compounds is determined by prevalent electrically active defects. The Seebeck effect measurements at room temperature reveal the $n$-type conductivity in the (In,Fe)Sb/GaAs structure and $p$-type in the (Ga,Fe)Sb/GaAs structure.

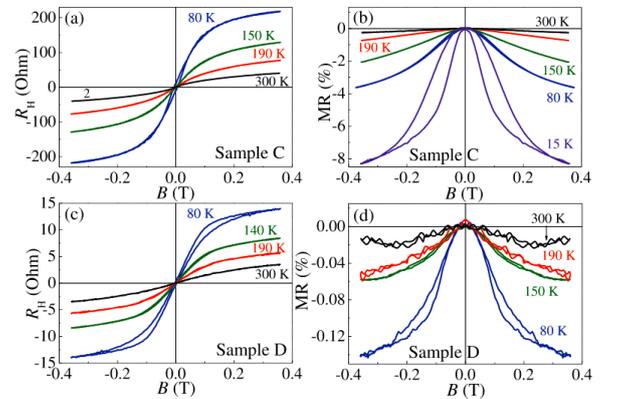

FIG. 11. $R_H(B)$ dependences (a, c) and MR curves (b, d) at various temperatures in the (In,Fe)Sb/GaAs (sample C) and (Ga,Fe)Sb/GaAs (sample D) structures. $B$ is applied perpendicular to the sample plane.

Figure 11(a, c) shows the dependences of the Hall resistance on the external magnetic field ($R_H(B)$) at various temperatures for the (In,Fe)Sb and (Ga,Fe)Sb layers. At all temperatures the $R_H(B)$ dependences are nonlinear with a saturation at $B \approx 0.2$ T, therefore, the anomalous Hall effect (AHE) for the obtained (In,Fe)Sb and (Ga,Fe)Sb layers is observed up to room temperature. The $R_H(B)$ dependences exhibit the p-type sign both for the n-type (In,Fe)Sb [6-9] and p-type (Ga,Fe)Sb [3-5] layers, i.e. the $R_H(B)$ curves are determined by the p-type AHE. Figure 11(b, d) shows the MR at various temperatures for the (In,Fe)Sb and (Ga,Fe)Sb layers for $B$ applied perpendicular to the sample plane. A negative MR is observed for both samples up to room temperature. Note that the MR value at a given temperature for the (In,Fe)Sb layer is much higher (more than an order of magnitude) than that for the (Ga,Fe)Sb layer [Figure 11(b, d)].

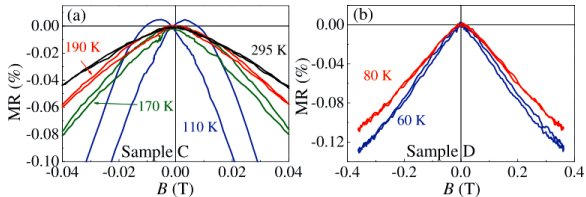

FIG. 12. (a) and (b) MR curves measured at various temperatures on the (In,Fe)Sb/GaAs (sample C) and (Ga,Fe)Sb/GaAs (sample D) structures, respectively. Magnetic field is applied in the sample plane.

In Ref. [6,9] we determined the Curie temperature ($T_C$) of the (In,Fe)Sb layers by the examination of the temperature evolution of MR curves for $B$ applied in the layer plane. In this case the MR curves are hysteretic due to the predominant in-plane orientation of the easy magnetization axis [6,9]. Figure 12(a) exhibits MR curves (with $B = \pm 0.04$ T lying in the sample plane) at selected temperatures for the (In,Fe)Sb layer. A clear hysteretic character of the MR curves is observed up to 190 K (Figure 12(a)). Hence, $T_C$ for the (In,Fe)Sb layer is about 190 K. This $T_C$ is close to the Curie point (170 K) for the (In,Fe)Sb layer with $Y_{Fe} = 0.17$ and $T_g = 200$ºC investigated in Ref. [6]. Figure 12(b) exhibits in-plane MR curves ($B = \pm 0.4$ T) at 60 K and 80 K for the (Ga,Fe)Sb layer. The MR value for the (Ga,Fe)Sb layer is much smaller, and the used method is not appropriate for the determination of $T_C$ in sample D.

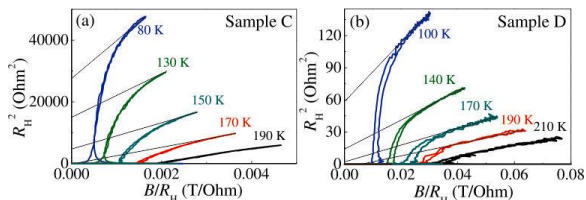

FIG. 13. (a) and (b) Arrot plots of $R_H(B)$ curves for the (In,Fe)Sb/GaAs (sample C) and (Ga,Fe)Sb/GaAs (sample D) structures, respectively.

Inasmuch as $R_H(B)$ curves are determined by the AHE [Figure 11 (a, c)], the Curie point can be estimated by Arrot plots of the $R_H(B)$ curves. Figure 13 shows the Arrot plots at selected temperatures for the (In,Fe)Sb and (Ga,Fe)Sb layers. The plots yield $T_C \approx 170$ K both for the (In,Fe)Sb and (Ga,Fe)Sb layers. Note that for the (In,Fe)Sb layer the $T_C$ value from the Arrot plots is lower than that determined from the analysis of the in-plane MR curves (Figure 12), probably due to the in-plane magnetic anisotropy. With a sufficiently high degree of confidence, we can conclude that the Curie temperature for the (Ga,Fe)Sb layer is equal to (at least not lower than) ≈ 170 K. This $T_C$ value is consistent with the Curie point of (Ga,Fe)Sb obtained by molecular beam epitaxy (MBE) at 250ºC and a Fe content of 8.5 at. % [4].

## IV. DISCUSSION

Our investigations reveal that the growth temperature is a critical parameter in the formation of InSb layers heavily doped with iron. For the used technique of pulsed laser sputtering and a Fe content of about 10 at. %, a $T_g$ increase from 220ºC to 300ºC leads to the coalescence of the Fe atoms and the formation of a secondary crystalline phase in the (In,Fe)Sb layer. Importantly, the (Ga,Fe)Sb compound is more resistant to the formation of the Fe-enriched secondary phase. The (Ga,Fe)Sb layer with a Fe content of about 10 at. % obtained at $T_g = 300$ºC, in contrast to the (In,Fe)Sb layer, is single-phase (Figure 2). The investigations also reveal that the (In,Fe)Sb and (Ga,Fe)Sb layers grown at $T_g = 220$ºC and with a Fe content of $10 \pm 1$ at. % are diluted magnetic semiconductors.

As stated above (Figure 7), hysteretic negative MR was observed in diode structure B at 10 K. A similar hysteretic MR curve was also observed at 15 K in the single (In,Fe)Sb layer [Figure 11(b)]. Taking into account the much higher values of the resistivity (Figure 10) and of the (negative) MR (Figure 11) for the (In,Fe)Sb layer than for the (Ga,Fe)Sb one, it can be concluded that the MR observed in diode structure B is determined by the current transfer through the (In,Fe)Sb layer. In structure B the (In,Fe)Sb layer acts as a relatively high series resistance with a high hysteretic negative MR. The observation of the AHE and negative MR reveals that the charge carrier transport in the (In,Fe)Sb and (Ga,Fe)Sb layers is spin-dependent. Consequently, under a forward bias of structure B, the (In,Fe)Sb layer emits spin-polarized electrons, while the (Ga,Fe)Sb layer emits spin-polarized holes. It can be expected that under reverse bias the flow of minority carriers from the DMS layers also is spin-polarized.

In the fabricated p-i-n diode structure B the (In,Fe)Sb and (Ga,Fe)Sb layers are ferromagnetic below ≈ 170 K. It is of interest to fabricate a heterostructure containing n-type (In,Fe)Sb and p-type (Ga,Fe)Sb layers with the Curie point above room temperature. In Ref. [9] we obtained smooth single-phase (In,Fe)Sb layers grown at 200ºC with a Fe content of about 12.5 at. % and $T_C$ close to room temperature. Similar growth conditions ($T_g = 200 -$ 220ºC and a Fe content in the range from 10.0 – 14.5 at. %) were used in the MBE deposition of (In,Fe)Sb layers with $T_C > 300$ K [7,8]. A growth temperature increase above 250ºC for the (In,Fe)Sb layers with a Fe content above 10 at. % leads to a coalescence of the Fe atoms into Fe clusters. Apparently, the optimal growth temperature for (In,Fe)Sb is about 200 – 220ºC. Inasmuch as the (Ga,Fe)Sb compound is resistant to the formation of Fe clusters, it is probable that for (Ga,Fe)Sb the optimal $T_g$ is higher than for (In,Fe)Sb. Our study reveals that for the single-phase (Ga,Fe)Sb compound $T_g$ can be increased at least up to 300ºC. In Ref. [5] we obtained ferromagnetic (Ga,Fe)Sb layers at $T_g = 350$ºC and 400ºC. For the MBE of (Ga,Fe)Sb layers, $T_g = 250$ºC was used [3,4]. Currently, (Ga,Fe)Sb layers with $T_C > 300$ K and $T_g$ below 250ºC have not yet been obtained, but this possibility cannot excluded.

Thus, it can be assumed that by choosing the optimal growth temperature and Fe content it is possible to grow in one technological cycle single-phase multilayer epitaxial structures based on the (In,Fe)Sb and (Ga,Fe)Sb DMS with the Curie point above room temperature.

## V. CONCLUSION

In summary, epitaxial *p-i-n* structures comprising *n*-type (In,Fe)Sb and *p*-type (Ga,Fe)Sb layers and separate constituent (In,Fe)Sb and (Ga,Fe)Sb layers were deposited using pulsed laser sputtering of InSb, GaAs, GaSb, Sb and Fe targets in a vacuum. TEM and EDS investigations revealed that the phase composition of the (In,Fe)Sb compound depends on the growth temperature ($T_g$). A $T_g$ increase from 220ºC to 300ºC leads to a coalescence of Fe atoms and a secondary crystalline phase formation in the (In,Fe)Sb layer with a total Fe content of ≈ 10 at. %. At the same time, the $Ga_{0.8}Fe_{0.2}Sb$ layers obtained at 220ºC and 300ºC are single-phase. The $In_{0.8}Fe_{0.2}Sb$ and $Ga_{0.8}Fe_{0.2}Sb$ layers grown at 220ºC are diluted magnetic semiconductors with $T_C \approx 190$ K and 170 K, respectively. The three-layer (In,Fe)Sb/GaAs/(Ga,Fe)Sb structure deposited on a $n^+$-GaAs substrate at 220ºC with a Fe content of $10 \pm 1$ at. % has a rather high crystalline quality, exhibits the spin-polarized charge carrier transport through DMS layers in the biased state, and can be considered as a prototype of a bipolar spintronic device based on Fe-doped III-V semiconductors.

**ACKNOWLEDGMENTS**
This study was supported by the Russian Science Foundation (grant № 18-79-10088).